\newcommand{\eat}[1]{}
\newcommand{\stitle}[1]{\noindent{\bf #1}}
\documentclass[sigconf,nonacm]{acmart}
\usepackage{tcolorbox}
\usepackage{lipsum}
\usepackage{enumitem}
\usepackage[utf8]{inputenc}
\usepackage{mdframed}
\AtBeginDocument{%
  }

\usepackage{multirow}
\setcopyright{acmlicensed}
\copyrightyear{2025}
\acmYear{2025}
\acmDOI{}
\acmConference[KDD Cup'25]{2025 KDD Cup Workshop for Multimodal Retrieval Augmented Generation}

\acmISBN{}

\begin{document}

\title{Solution for Meta KDD Cup' 25: A Comprehensive Three-Step Framework for Vision Question Answering}

\author{Zijian Zhang}
\authornote{All authors contributed equally to this research.}
\affiliation{%
  \institution{MeiTuan}
  \city{Shanghai}
  \country{China}
}
\email{zhangzijian14@meituan.com}

\author{Xiaocheng Zhang}
\authornotemark[1]
\affiliation{%
  \institution{MeiTuan}
  \city{BeiJing}
  \country{China}}
\email{zhangxiaocheng@meituan.com}

\author{Yang Zhou}
\authornotemark[1]
\affiliation{%
  \institution{MeiTuan}
  \city{ShangHai}
  \country{China}
}
\email{zhouyang96@meituan.com}

\author{Zhimin Lin}
\authornotemark[1]
\affiliation{%
 \institution{MeiTuan}
 \city{BeiJing}
 \country{China}}
\email{linzhimin@meituan.com}

\author{Peng Yan}
\authornote{Corresponding Author}
\affiliation{%
  \institution{MeiTuan}
  \city{BeiJing}
  \country{China}}
\email{yanpeng04@meituan.com}

\begin{abstract}
Vision Large Language Models (VLLMs) have improved multi-modal understanding and visual question answering (VQA), but still suffer from hallucinated answers. Multi-modal Retrieval-Augmented Generation (RAG) helps address these issues by incorporating external information, yet challenges remain in visual context comprehension, multi-source retrieval, and multi-turn interactions.
To address these challenges, Meta constructed the CRAG-MM benchmark and launched the CRAG-MM Challenge at KDD Cup 2025, which consists of three tasks.
This paper describes the solutions of all tasks in Meta KDD Cup’25 from \textbf{BlackPearl} team. 
We use a single model for each task, with key methods including data augmentation, RAG, reranking, and multi-task fine-tuning. Our solution achieve automatic evaluation rankings of 3rd, 3rd, and 1st on the three tasks, and win second place in Task3 after human evaluation.
\end{abstract}



\maketitle

\section{Introduction}

Vision Large Language Models (VLLMs) have made significant progress in enabling multi-modal understanding and visual question answering (VQA). However, they still struggle with generating hallucinated answers and handling complex or long-tail queries that require abilities such as recognition, OCR, and knowledge integration\cite{qiu2024snapntell,yu2023mm}.
The Retrieval-Augmented Generation (RAG) paradigm extends to multi-modal (MM) input and shows promise in overcoming VLLM's knowledge limitations. Given an image and a question, an MM-RAG system generates a search query, retrieves relevant external information, and provides grounded answers\cite{gao2023retrieval}.
Despite its potential, MM-RAG still faces challenges in understanding visual context, retrieving relevant information, integrating multi-source data, and supporting multi-turn conversations.

To address these issues, Meta introduces CRAG-MM with the aim of reliably evaluating MM-RAG QA systems.
CRAG-MM is a visual question-answering benchmark focused on factual questions, featuring 5,000 diverse images—including 3,000 egocentric photos from RayBan Meta smart glasses—across 13 domains. It includes four types of questions, from simple image-based queries to complex ones requiring multi-source retrieval and reasoning, as well as both single-turn and multi-turn conversations for comprehensive evaluation of MM-RAG solutions.

Based on this benchmark, the CRAG-MM Challenge is the sole event in the 2025 KDD Cup, aiming to encourage the development and evaluation of advanced MM-RAG systems.
Meta designed three competition tasks. Task1 and Task2 contain single-turn questions, where the former provides image-KG-based retrieval, and the latter additionally introduces web retrieval; Task3 focuses on multi-turn conversations:
\begin{enumerate}
    \item \textbf{Task1: Single-source Augmentation.} Only an image-based mock KG is provided to test the basic answer generation capability of MM-RAG systems.
    \item \textbf{Task2: Multi-source Augmentation.} An additional web search mock API is provided to test how well the MM-RAG system synthesizes information from different sources.
    \item \textbf{Task3: Multi-turn QA.} To test context understanding for smooth multi-turn conversations.
\end{enumerate}
The solution of each team must be submitted for inference online, with each generated answer having to be produced in 30 seconds and restricted to the use of the Llama model.

We form the BlackPearl team and participate in all three tasks, achieving automatic evaluation rankings of 3rd, 3rd, and 1st, respectively. After human evaluation, we secure 2nd place in Task3. This paper provides a detailed description of our solutions for all three tasks, with major improvements including data augmentation, RAG, reranking, and multi-task fine-tuning.
Our code is available on github~\footnote{https://github.com/BlackPearl-Lab/KddCup-2025-CRAG-MM-Solution}.
\section{Solution to Task1}
\label{sec:1}

\begin{figure}[t]
\centering
\includegraphics[width=0.5\textwidth]{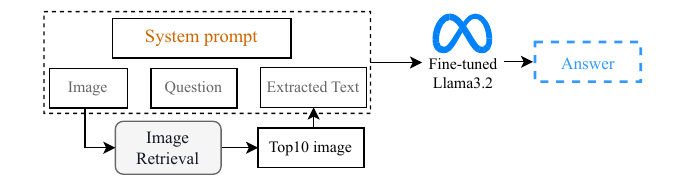}
\caption{Illustration of Task1 framework.}
\label{fig:task1_framework}
\end{figure}

This section presents our solution for Task1, including key components such as image retrieval, data augmentation, and model fine-tuning. Some of these techniques are also applied to Task2 and Task3. Figure \ref{fig:task1_framework} illustrates our inference framework for Task1.

\subsection{Image Retrieval}
We utilize the official retrieval tool for image retrieval, representing the image as a vector, and retrieving the top 10 images most similar to the current input image. Instead of directly adding the retrieved images to the model input, we first extract the textual information from the structured data associated with these images. This textual information is then incorporated into the model input as additional context, thereby enhancing the model's response quality.

\subsection{Data augmentation}
\label{sec:data_augmentation}
\begin{figure}[t]
\centering
\includegraphics[width=0.5\textwidth]{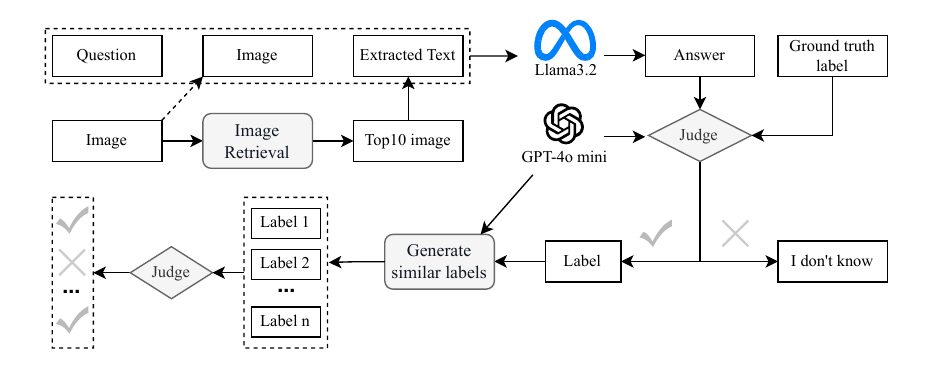}
\caption{Data augmentation method for Task1.}
\label{fig:task1}
\end{figure}

This subsection introduces our data augmentation(DA) approach for Task1, with the overall workflow illustrated in Figure \ref{fig:task1}. For each sample, we first use the image retrieval module described in the previous subsection to obtain relevant information. Then, the original question, image, and retrieved information are fed into Llama3.2 for inference, resulting in an initial answer from the model. Next, we use GPT-4o mini to verify the answer against the ground truth label. If the answer is identified as a hallucination, the label for this sample is set to “I don't know”; otherwise, the sample is retained for the next stage. Specifically, we use GPT-4o mini to generate $n$ similar labels ($n$=10) for the verified label. All generated labels are then re-verified to filter out hallucinated labels. The remaining $m$ labels, together with the original question and image, are used to construct m additional samples, which are utilized to enhance the training process.

\subsection{Model Fine-Tuning and Inference}
\stitle{Base Model.} We follow the contest instructions to use the LLama series LLM~\footnote{https://llama.meta.com/}. Considering the limited running time, we use the Llama-3.2-11B-Vision-Instruct model as the base model.

\stitle{Fine-Tuning Data.}
We randomly split the original Task1 dataset into training and validation sets at an 8:2 ratio based on images. The data augmentation methods described in Sec.\ref{sec:data_augmentation} are applied to the training set to construct the fine-tuning data.

\stitle{Fine-Tuning.}
Due to the high requirements for memory efficiency and training speed during the competition, parameter-efficient fine-tuning methods are more suitable. Therefore, we adopted LoRA (Low-Rank Adaptation)\cite{hu2022lora}, which enables efficient fine-tuning with only a small number of additional parameters, thus saving resources. More details on the fine-tuning parameters can be found in Sec.\ref{sec:experiment_settings}.

\stitle{Inference.}
To meet the inference time constraints of the competition, we utilized vLLM\cite{kwon2023efficient} for model inference. vLLM offers significantly higher throughput and lower latency compared to standard inference frameworks, making it well-suited for efficient large-scale deployment. The overall inference process is shown in Figure \ref{fig:task1_framework}. First, we retrieve the top 10 most similar images and extract only their associated textual information. This textual information is then combined with the original input and fed into the model for inference.

\section{Solution to Task2 and Task3}
\begin{figure*}[t]
\centering
\includegraphics[width=0.95\textwidth]{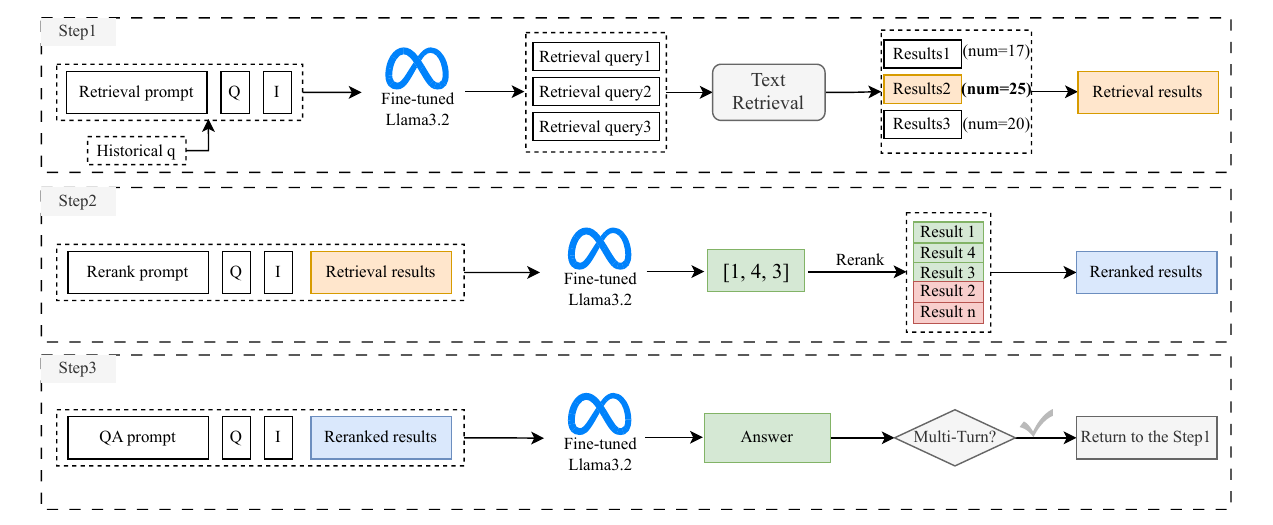}
\caption{Illustration of Task2 and Task3 framework. The entire process consists of three steps. 'Q' and 'I' represent Question and Image, respectively.}
\label{fig:task2_3}
\end{figure*}

This section presents our solution for Task2 and Task3. The framework is illustrated in Figure \ref{fig:task2_3}. The key components of our solution include web retrieval, reranking, and multi-task fine-tuning.

\subsection{Web retrieval}
Step1 in Figure \ref{fig:task2_3} illustrates our web retrieval process. First, we concatenate the retrieval prompt, image, question, and historical questions to construct the input for the Llama model. This input is then fed into the model for inference to generate a query for retrieval. This process is repeated several times with randomization to obtain multiple queries. For each query, we use the official tool to perform a retrieval and obtain multiple groups of results. Finally, we retain the group with the largest number of results as the final retrieval result.

\subsection{Reranking}

Due to limitations on input length and online inference time, increasing the proportion of high-quality results in the retrieval output becomes crucial. To address this, we designed a reranking module to prioritize high-quality results as much as possible. Notably, the winning solution of the 2024 KDD Cup also adopted a reranking approach\cite{xia2024winning}.
To better accommodate the data formats of other tasks and facilitate more convenient training, we did not adopt the top-ranking solutions from related Kaggle competitions\cite{eedi-mining-misconceptions-in-mathematics,wsdm-cup-multilingual-chatbot-arena}.
We developed a listwise reranking method, as illustrated in Step2 of Figure \ref{fig:task2_3}. 
Specifically, we concatenate the reranking prompt, question, image, and the retrieval results obtained from Step1 as the model input, where each retrieval result is labeled with identifiers such as '1', '2', and '3'.
This input is then fed into the model for inference.
The model should output the most relevant info number list in the format $[x, xx, xxx, ...]$. If there are no relevant items, output $[]$.

\subsection{Multi-Task Fine-Tuning and Inference}
Since vLLM does not support loading mLLama's LoRA weights, we adopt multi-task fine-tuning to enable the unified model to adapt to various task formats.
The prompts corresponding to each task are provided in Appendix \ref{appendix_A}.

\stitle{Base Model.}
As in Task 1, we choose Llama-3.2-11B-Vision-Instruct as the base model.

\stitle{Fine-Tuning Data.} 
We randomly split the original Task2 and Task3 datasets into training and validation sets at an 8:2 ratio based on images. Each multi-turn dialogue session is divided into multiple samples. All fine-tuning data consists of three parts: retrieval query generation task data, re-ranking task data, and question answering task data.

For the \textbf{retrieval query generation task data}, we use the retrieval prompt combined with the historical questions (for multi-turn scenarios), the current question, and the image to construct the input, which is then fed into the original Llama model to generate the query. This generated query serves as the label required for fine-tuning in this task.

For the \textbf{reranking task data}, We take the question, image, and one of the retrieval results as input and feed them into GPT-4o. GPT-4o then determines whether the retrieval result is helpful for answering the question and outputs either True or False. The detailed prompt is shown in Appendix \ref{appendix_A_rdc_prompt}.

For the \textbf{question answering task data}, similar to Task1, we use the QA prompt combined with the question, image, historical questions, and the retrieved results as input to Llama for answer generation. Subsequently, we use GPT-4o mini to determine the consistency between the ground truth answer and the predicted answer. If they are inconsistent, the fine-tuning label for that sample is changed to “I don't know.” If they are consistent, we use the model-generated answer as the fine-tuning label instead of the ground truth. This approach allows the model to focus more on learning the task pattern itself rather than the more challenging transfer of specific text styles.

\stitle{Fine-Tuning.}
Similar to Task1, we use LoRA for model fine-tuning. Detailed parameter settings can be found in Section \ref{sec:experiment_settings}.

\stitle{Inference.}
Figure \ref{fig:task2_3} illustrates the overall inference process, where a unified model is used to perform multiple tasks. For each sample, multiple diverse queries are first generated for retrieval. The retrieval tool is then called with these queries to obtain multiple groups of results, and the group with the largest number of results is retained. This group is further reranked, and up to the top 10 results are kept. These results are then used as additional information and fed into the model for inference to obtain the answer. For Task2, the process ends here; for Task3, the workflow proceeds to the next round of answering, returning to Step1 to repeat the process. The inference acceleration framework also utilizes vLLM.

\section{Experiments}
In this section, we present our main results and ablation studies for some crucial components.

\subsection{Experiment Settings.}
\label{sec:experiment_settings}
\stitle{Metrics}
This competition adopts exactly the same metrics and methods used in the CRAG\cite{yang2024crag} competition to assess the performance of MM-RAG systems. For each question in the evaluation set, the answer is scored as:
\begin{itemize}
    \item \textit{Perfect} (fully correct) → Score: 1.0
    \item \textit{Acceptable} (useful but with minor non-harmful errors) → Score: 0.5
    \item \textit{Missing} (e.g., “I don’t know”, “I’m sorry I can’t find …”) → Score: 0.0
    \item \textit{Incorrect} (wrong or irrelevant answer) → Score: -1.0
    \item \textit{Truthfulness Score}: The average score across all examples in the evaluation set for a given MM-RAG system.
\end{itemize}
For multi-turn conversations, the evaluation is terminated if two consecutive answers are incorrect, and all remaining turns in the conversation are assigned a score of zero\cite{bai2024mt}. The final result is the average score across all multi-turn conversations.

\stitle{Parameter Settings}
Our implementations are based on Pytorch.
For Task1, the number of training epochs and the learning rate are set to 2 and 5e-5, respectively.
For Task2 and Task3, the number of training epochs and the learning rate are set to 10 and 5e-6, respectively.
The rank, alpha, and dropout parameters of LoRA are set to 64, 128, and 0.05, respectively.
The warmup ratio is set to 0.03.
For all tasks, during the inference phase, the maximum input length is set to 8192, and the maximum output length is set to 75. The temperature in vLLM is set to 0.0 for all cases, except when sampling retrieval queries, where it is set to 0.8.
The maximum number of retrieval results for Task 2 and Task 3 is set to 30.
Fine-tuning is performed on $8\times A100$ GPUs.

\subsection{Overall Performance}

Tables \ref{tab:task1_leaderboard}, \ref{tab:task2_leaderboard}, and \ref{tab:task3_leaderboard} show the leaderboard results of our solution for each task, all evaluated automatically.
We designed a dedicated solution for Task1, achieving a score of 0.043 and ranking third on the leaderboard.
The overall framework for Task2 and Task3 is the same, with the difference being that the input for Task3 includes historical information. Our solution achieved scores of 0.107 and 0.175 for Task2 and Task3, ranking third and first, respectively.

Since the automatic evaluation of VLLMs can be somewhat uncertain, the organizers conducted a human evaluation for the top 10 teams, correcting test samples that were judged incorrect by the automatic evaluation but actually should be considered correct. As a result, the scores from human evaluation are generally higher than those from automatic evaluation.
Because our solution has a relatively low hallucination rate, the improvement from human evaluation was smaller compared to other teams. Therefore, our final ranking was slightly lower than the automatic leaderboard. Table \ref{tab:final_score} presents the final scores and rankings after human evaluation, where our solution ranked second in Task3 with a score of 30.9\%.

\begin{table}[t]
\renewcommand{\arraystretch}{1.2}
\caption{The results of different evaluation metrics for the top 6 teams on the Task 1 leaderboard.}
\begin{tabular}{lcccc}
\toprule
Teams              & Score & M & H & A \\ \midrule
Dianping-Trust-Safety   & 0.073 & 0.711	   & 0.108	         & 0.181	    \\
db3                & 0.053	 & 0.801   & 0.073	         & 0.126	    \\
BlackPearl(Our)    & 0.043 & 0.897	   & 0.030         & 0.073   \\
y3h2     & 0.036 & 0.860	  & 0.052        & 0.088    \\
USTGZ-KIMI    & 0.029 & 0.940	   & 0.015	         & 0.044	    \\
Team\_NVIDIA    & 0.026 & 0.920	   & 0.027         & 0.053    \\
\multicolumn{5}{c}{......}                                      \\ \bottomrule
\end{tabular}
\label{tab:task1_leaderboard}
\end{table}
\begin{table}[t]
\renewcommand{\arraystretch}{1.2}
\caption{The results of different evaluation metrics for the top 6 teams on the Task 2 leaderboard.}
\begin{tabular}{lcccc}
\toprule
Teams              & Score & M & H & A \\ \midrule
db3                & 0.124	 & 0.688   & 0.094		         & 0.218	    \\
Dianping-Trust-Safety   & 0.119	 & 0.625	   & 0.128		         & 0.247		    \\
BlackPearl(Our)    & 0.107	 & 0.723		   & 0.085	         & 0.192	   \\
Team\_NVIDIA    & 0.075 & 0.774		   & 0.075	         & 0.151	    \\
AcroYAMALEX     & 0.057 & 0.665		  & 0.139	        & 0.196	    \\
zmf    & 0.051	 & 0.789	   & 0.080		         & 0.131	    \\
\multicolumn{5}{c}{......}                                      \\ \bottomrule
\end{tabular}
\label{tab:task2_leaderboard}
\end{table}
\begin{table}[t]
\renewcommand{\arraystretch}{1.2}
\caption{The results of different evaluation metrics for the top 6 teams on the Task 3 leaderboard.}
\begin{tabular}{lcccc}
\toprule
Teams              & Score & M & H & A \\ \midrule
BlackPearl(Our)    & 0.175	 & 0.638		   & 0.094	         & 0.269	   \\
db3                & 0.172		 & 0.533	   & 0.147	       & 0.319	    \\
Dianping-Trust-Safety   & 0.121 & 0.706		   & 0.086	         & 0.208	    \\
Team\_NVIDIA    & 0.119 & 0.713		   & 0.084	         & 0.203	    \\
y3h2     & 0.104 & 0.827		  & 0.035	        & 0.138	    \\
AcroYAMALEX    & 0.100 & 0.679		   & 0.111			         & 0.211		    \\
\multicolumn{5}{c}{......}                                      \\ \bottomrule
\end{tabular}
\label{tab:task3_leaderboard}
\end{table}
\begin{table}[t]
\renewcommand{\arraystretch}{1.2}
\caption{Final evaluation process and team scores}
\begin{tabular}{llc}
\toprule
Task                   & Team                  & Score \\ \midrule
\multirow{3}{*}{Task1} & Dianping-Trust-Safety & 12.8  \\
                       & db3                   & 8.4   \\
                       & cruise                & 6.7   \\ \midrule
\multirow{3}{*}{Task2} & Team\_NVIDIA          & 23.3  \\
                       & db3                   & 22.1  \\
                       & AcroYAMALEX           & 21.4  \\ \midrule
\multirow{3}{*}{Task3} & db3                   & 36.8  \\
                       & BlackPearl(Our)       & 30.9  \\
                       & Dianping-Trust-Safety & 29.7  \\ \bottomrule
\end{tabular}
\label{tab:final_score}
\end{table}

\begin{table}[t]
\renewcommand{\arraystretch}{1.2}
\caption{Representative experimental results from the Task1 leaderboard.}
\begin{tabular}{lc}
\toprule
Methods                & Score(\%) \\ \midrule
Fine-tuning            & 0.01  \\
Fine-tuning(\textit{+RAG})      & 0.02  \\
Fine-tuning(\textit{+RAG, +DA}) & 0.043 \\ \bottomrule
\end{tabular}
\label{tab:task1_representative}
\end{table}


\subsection{Representative Experimental Results}
We selected several representative experimental results to more clearly demonstrate the effectiveness of our approach.

Table \ref{tab:task1_representative} presents some experimental results for Task1. The fine-tuned model without retrieval achieved a score of only 0.01. After adding image retrieval, the score increased to 0.02. With data augmentation to enrich the fine-tuning data, the score further improved to 0.043.

Table \ref{tab:task3_representative} shows some experimental results for Task3 as an example. The score of the model fine-tuned with original RAG data was 0.07. After adding refusal data, the score increased to 0.1322. We then sampled multiple queries for retrieval to expand the retrieval results, raising the score to 0.1471. Incorporating reranking to increase the proportion of high-quality data in the input further improved the score to 0.1505. Finally, slightly reducing the proportion of refusal data in the training set led to a score of 0.1755, ranking first in the automatic evaluation. These methods significantly improved the scores, demonstrating their effectiveness.


\begin{table}[t]
\renewcommand{\arraystretch}{1.2}
\caption{Representative experimental results from the Task3 leaderboard.}
\begin{tabular}{lc}
\toprule
Methods                                           & Score(\%)                  \\ \midrule
Fine-tuning(\textit{RAG})                         & 0.0700                     \\
\textit{+Refusal Data Construction}               & 0.1322                     \\
\textit{+Multi-query Retrieval}                   & 0.1471                     \\
\textit{+Reranking}                               & 0.1505 \\
\textit{+Reducing the Proportion of Refusal Data} & 0.1755 \\ \bottomrule
\end{tabular}
\label{tab:task3_representative}
\end{table}

\section{Conclusion}
The Meta CRAG-MM Challenge is the first MM-RAG competition for the KDD Cup and serves as an important driver for the development of VLLMs and VQA. We have presented our approaches to all three tasks in the contest. Due to differences in retrieval sources, we developed distinct solutions for Task 1 and Task 2/3, each with its own focus. In Task 1, we proposed a novel data augmentation strategy, while in Task 2 and Task 3, we adopted diversified retrieval, re-ranking, and multi-task fine-tuning to enhance performance. As a result, our solution achieved automatic evaluation rankings of 3rd, 3rd, and 1st in the three tasks, and won second place in Task 3 after human evaluation.

\label{sec:conclusion}

\bibliographystyle{ACM-Reference-Format}

\appendix
\section{Prompts Used in the Competition}
\label{appendix_A}

\subsection{VQA Prompt for Task1}
\begin{figure*}[t]
\centering
\includegraphics[width=0.95\textwidth]{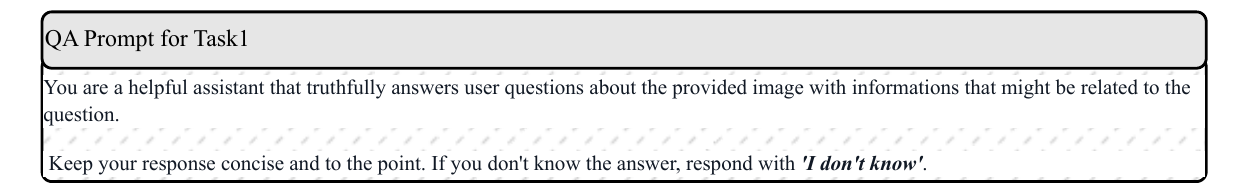}
\caption{QA prompt for Task1.}
\label{fig:task1_qa_prompt}
\end{figure*}

Figure \ref{fig:task1_qa_prompt} shows the VQA prompt for Task1.

\subsection{Data Augmentation Prompt for Task1}
\begin{figure*}[t]
\centering
\includegraphics[width=0.95\textwidth]{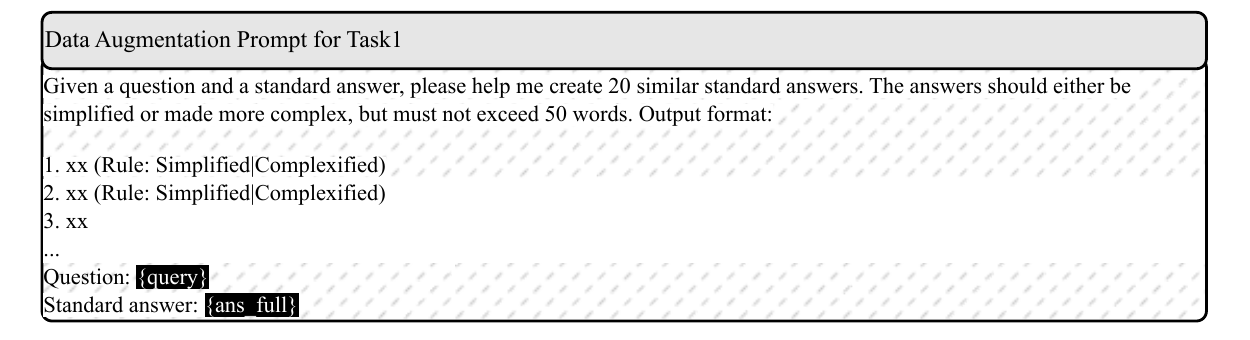}
\caption{Data augmentation prompt for Task1.}
\label{fig:task1_da_prompt}
\end{figure*}

As shown in the prompt in Figure \ref{fig:task1_da_prompt}, we instruct the model to generate diverse answers along both simple and complex dimensions while preserving the original meaning.

\subsection{Retrieval Query Generation Prompt for Task2 and Task3}
\begin{figure*}[t]
\centering
\includegraphics[width=0.95\textwidth]{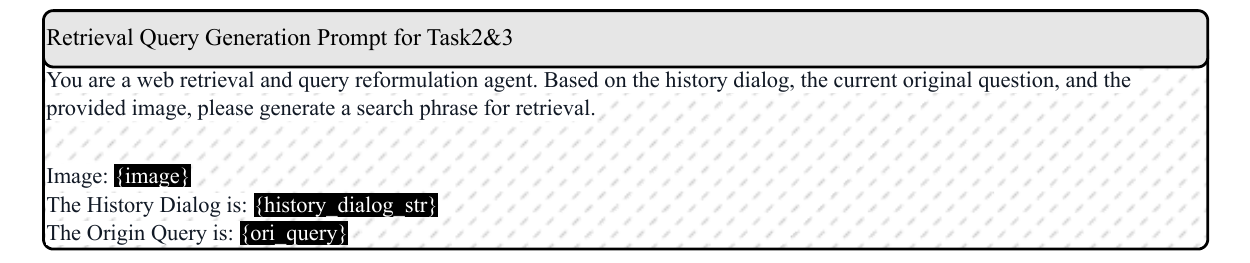}
\caption{Retrieval query generation prompt for Task2/3.}
\label{fig:task2_3_rq_prompt}
\end{figure*}

Figure \ref{fig:task2_3_rq_prompt} shows the retrieval query generation prompt for Task2 and Task3.

\subsection{Prompt for Reranking Data Construction in Task2 and Task3}
\label{appendix_A_rdc_prompt}
\begin{figure*}[t]
\centering
\includegraphics[width=0.95\textwidth]{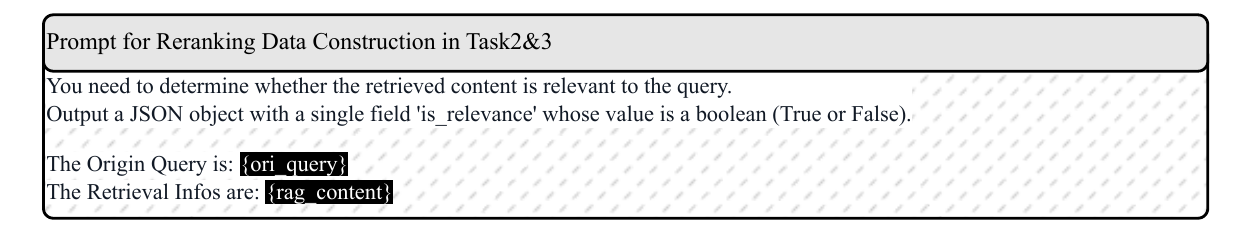}
\caption{Prompt for reranking data construction in Task2/3.}
\label{fig:task2_3_rdc_prompt}
\end{figure*}

Figure \ref{fig:task2_3_rdc_prompt} shows the prompt we used with a large model to construct fine-tuning data for the reranking task.

\subsection{Rerank Prompt for Task2 and Task3}
\begin{figure*}[t]
\centering
\includegraphics[width=0.95\textwidth]{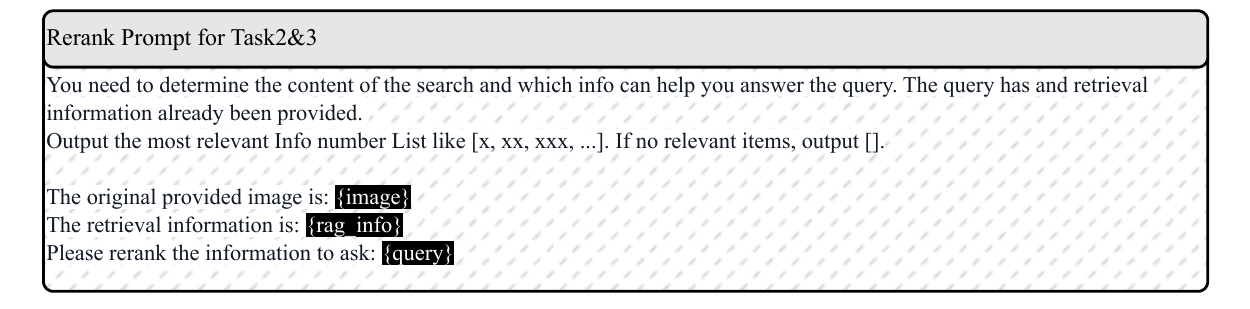}
\caption{Rerank prompt for Task2/3.}
\label{fig:task2_3_rerank_prompt}
\end{figure*}

Figure \ref{fig:task2_3_rerank_prompt} shows the reranking prompt.

\subsection{VQA Prompt for Task2 and Task3}
\begin{figure*}[t]
\centering
\includegraphics[width=0.95\textwidth]{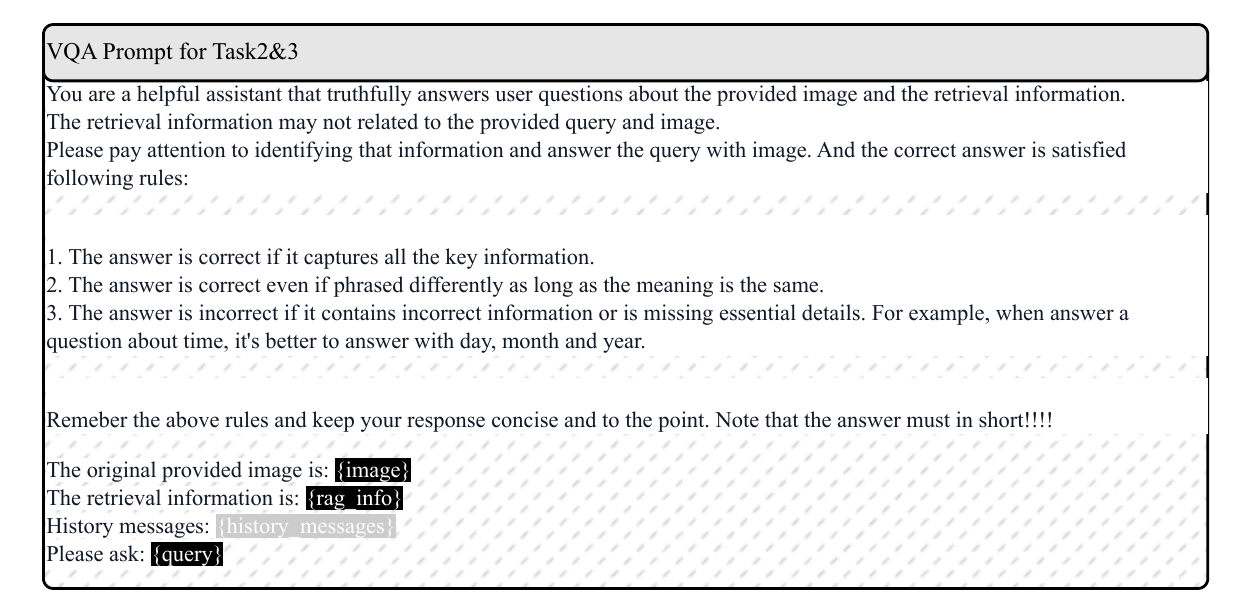}
\caption{VQA prompt for Task2/3.}
\label{fig:task2_3_vqa_prompt}
\end{figure*}

Figure \ref{fig:task2_3_vqa_prompt} shows the prompt used for generating the final answer after obtaining retrieval results in Task2 and Task3.

\end{document}